\documentclass[useAMS,usenatbib]{mn2e}                                  
\usepackage{graphicx, natbib, amssymb, color}

\title[Diffuse gamma-ray emission in Coma]{Constraints on diffuse
gamma-ray emission from structure formation processes in the Coma cluster}
\author[Zandanel \& Ando]{Fabio Zandanel$^{*}$ and Shin'ichiro
Ando$^{\dagger}$\\% 
GRAPPA Institute, University of Amsterdam, 1098XH Amsterdam, Netherlands}
\begin{document}

\date{\today}
%\date{Accepted XXX. Received XXX; in original from XXX}

\pagerange{\pageref{firstpage}--\pageref{lastpage}} \pubyear{2013}

\maketitle		

\label{firstpage}

%%%%%%%%%%%%%%%%%%%%%%%%%%%%%%%%%%%%%%%%%%%%%%%%%%%%%%%%%%%%%%%%%%%
%%%%%%%%%%%%%%%%%%%%%%%%%%%%%%%%%%%%%%%%%%%%%%%%%%%%%%%%%%%%%%%%%%%
\begin{abstract}
We analyze 5-year (63 months) data of the Large Area Telescope on board {\it Fermi} 
satellite from the Coma galaxy cluster in the energy range between 100~MeV 
and 100~GeV. The likelihood analyses are performed with several templates
motivated by models predicting gamma-ray emission due to structure formation processes. 
We find no excess emission and derive the most stringent constraints 
to date on the Coma cluster above 100~MeV, and on the tested scenarios in general.
The upper limits on the integral flux range from $10^{-10}$ to $10^{-9}$~cm$^{-2}$~s$^{-1}$, 
and are stringent enough to challenge different scenarios. We find that the 
acceleration efficiency of cosmic ray protons and electrons at shocks must be 
below approximately 15\% and 1\%, respectively. Additionally, we argue that
the proton acceleration efficiency should be lower than 5\% in order to be 
consistent with radio data. This, however, relays on magnetic field estimates
in the cluster. In particular, this implies that the contribution to the diffuse extragalactic 
gamma-ray background due to gamma-rays from structure formation 
processes in clusters of galaxies is negligible, below 1\%. Finally, we discuss 
future detectability prospects for Astro-H, \emph{Fermi} after 10-yr of operation, and 
the Cherenkov Telescope Array.
\end{abstract}


%%%%%%%%%%%%%%%%%%%%%%%%%%%%%%%%%%%%%%%%%%%%%%%%%%%%%%%%%%%%%%%%%%%
%%%%%%%%%%%%%%%%%%%%%%%%%%%%%%%%%%%%%%%%%%%%%%%%%%%%%%%%%%%%%%%%%%%
\section{Introduction}
\label{sec:1}

\begingroup
\let\thefootnote\relax\footnotetext{* f.zandanel@uva.nl}
\let\thefootnote\relax\footnotetext{$\dagger$ s.ando@uva.nl}
\endgroup
 
According to the hierarchical scenario, structures form via accretion and merger
of smaller objects into larger ones. Clusters of galaxies are the latest and 
largest structures to have formed in the Universe. They have typical radii of a
few Mpc, and masses of about $10^{14}$--$10^{15} M_\odot$.
Dark matter contributes for about 80\% of their mass, gas for 15\% 
and galaxies for 5\%  \citep{2005RvMP...77..207V}. During the course of a 
cluster formation, part of its gravitational binding energy, on the order of $10^{61}$--$10^{63}$~erg, 
should be dissipated through turbulence and structure-formation (accretion and merger) shocks 
that accelerate charged particles such as protons and electrons. Even if only a small fraction 
of this energy goes into the particle acceleration, the process should be strong enough to make 
the clusters visible with non-thermal emissions such as radio synchrotron emission, and potentially 
in gamma-ray frequencies.

Diffuse radio synchrotron emission is detected in the Coma cluster both
as a central radio {\it halo} and as a peripheral radio {\it relic}
\citep[e.g.,][]{1997A&A...321...55D, 2011MNRAS.412....2B}.
This proves the presence of relativistic electrons and magnetic fields
permeating the intra-cluster medium (ICM).
Radio relics are apparently connected to structure-formation shocks
\citep[e.g.,][]{2011A&A...533A..35V}, even though the details of the
particle acceleration process at place are not clear yet
\citep[e.g.,][]{2013MNRAS.433..812O,pinzke13}.
Radio halos can be further divided into two categories: mini-halos and
giant halos. The former is associated with relaxed, cool-core clusters,
and typically extend over a few hundred kpc. The latter, such as the one
found in Coma, is typically associated with cluster mergers 
and have Mpc-sizes \citep[see][for a review]{2012A&ARv..20...54F}.
The generation mechanism of radio halos is currently debated between
re-acceleration (e.g., \citealp{2007MNRAS.378..245B,
2010arXiv1008.0184B, 2012arXiv1207.3025B}) and hadronic models
(e.g., \citealp{2004A&A...413...17P,2004MNRAS.352...76P,2008MNRAS.385.1211P,
2011A&A...527A..99E, 2012arXiv1207.6410Z}).
 
Cosmic ray (CR) electrons can generate X-ray and gamma-ray emission 
through inverse Compton (IC) up-scattering of cosmic-microwave-background 
(CMB) photons. 
Additionally, if the radio emission is due to secondary electrons
generated in hadronic interactions between CR protons and the ICM, it should 
be accompanied by a detectable gamma-ray emission 
generated from neutral pion decays.
 
Observationally, gamma-ray emission from clusters of galaxies were
searched for the last several years to decade, but all these attempts resulted in no
detection (for space-based cluster observations in the GeV-band, see
\citealp{2003ApJ...588..155R, 2010JCAP...05..025A, 2010ApJ...717L..71A,
2012AAS...21920701Z, 2012arXiv1207.6749H, 2012JCAP...07..017A,
2013arXiv1308.6278H}; for ground-based observations in the energy band 
above $\sim$100~GeV, see \citealp{2006ApJ...644..148P, 2008AIPC.1085..569P,
2009arXiv0907.0727T, 2009A&A...495...27A, 2009arXiv0907.3001D,
2009arXiv0907.5000G,cangaroo_clusters, 2009ApJ...706L.275A,
2010ApJ...710..634A,2011arXiv1111.5544M, 2012...VERITAS,
2012A&A...545A.103H}). Recently, both \cite{2013arXiv1308.5654T}
and \cite{2013arXiv1309.0197P} performed a joint likelihood analysis 
of about 50 galaxy clusters. They found a significant gamma-ray excess
in the direction of few objects, particularly Abell~400, which however is
interpreted either as coming from active galactic nuclei (AGN) within
the cluster or as un-modeled background.  

The Large Area Telescope (LAT) on board {\it Fermi}
satellite\footnote{http://fermi.gsfc.nasa.gov/} continuously surveys the
gamma-ray sky since August 2008, and its data represent a unique tool to
test diffuse CR emission models in clusters of galaxies.
Although many previous analyses of the {\it Fermi}-LAT data of
galaxy clusters were based on the assumption that the clusters were
simply point sources or have a very simple extended profile, a possible
cluster gamma-ray emission could have very different spatial structure 
compared with what were investigated so far.
Predicted gamma-ray emission profiles and spectra differ depending on
models of cluster formation as well as particle acceleration, and thus,
by detecting or constraining them, one could learn important physics
thereof, which is still completely missing and well awaited.

In this paper, we take a deeper look at the \emph{Fermi}-LAT data for
GeV gamma-ray emission from the Coma cluster and its possible diffuse
emission induced by CR interactions.
We chose Coma because it is one of the best studied clusters, which is
located in a local volume (its distance is about 100~Mpc).
It also shows evidence of recent dynamical activities such as particle
accelerations, as seen from presence of a giant radio halo
and a radio relic.
Together with the fact that there are no AGN found in the LAT data,
these make Coma an ideal environment to test CR-induced gamma-ray
emission.
We perform dedicated analyses of 63-month {\it Fermi}-LAT data using
well-motivated models for spatial emission distribution.
Besides the simplest point-source model, we investigate (1) models
based on hydrodynamical simulations of the cluster formation that also
trace interactions of CR protons with the ICM; (2) a model motivated by 
spatial profile of the radio relic, and (3) disk and ring-like profiles
motivated by scenarios where primary electrons accelerated by 
accretion shocks dominates the cluster high-energy emission.
We find no positive signatures in any of these scenarios, and thus, put
the most constraining upper limits to date on the gamma-ray flux from Coma 
for each model and interpret them in terms of constraints on parameters of 
cluster formation and CR physics.

This paper is organized as follows.
In Section~\ref{sec:2}, we describe details of the {\it Fermi}-LAT data
analyses for Coma.
Several theoretical model templates of the gamma-ray emission are
explained in Section~\ref{sec:3}.
We present our results in Section~\ref{sec:4} and conclude in
Section~\ref{sec:6}.

%%%%%%%%%%%%%%%%%%%%%%%%%%%%%%%%%%%%%%%%%%%%%%%%%%%%%%%%%%%%%%%%%%%
%%%%%%%%%%%%%%%%%%%%%%%%%%%%%%%%%%%%%%%%%%%%%%%%%%%%%%%%%%%%%%%%%%%
\section{Data Analysis}
\label{sec:2}

\begin{figure*}
\centering 
\includegraphics[width=.33\textwidth]{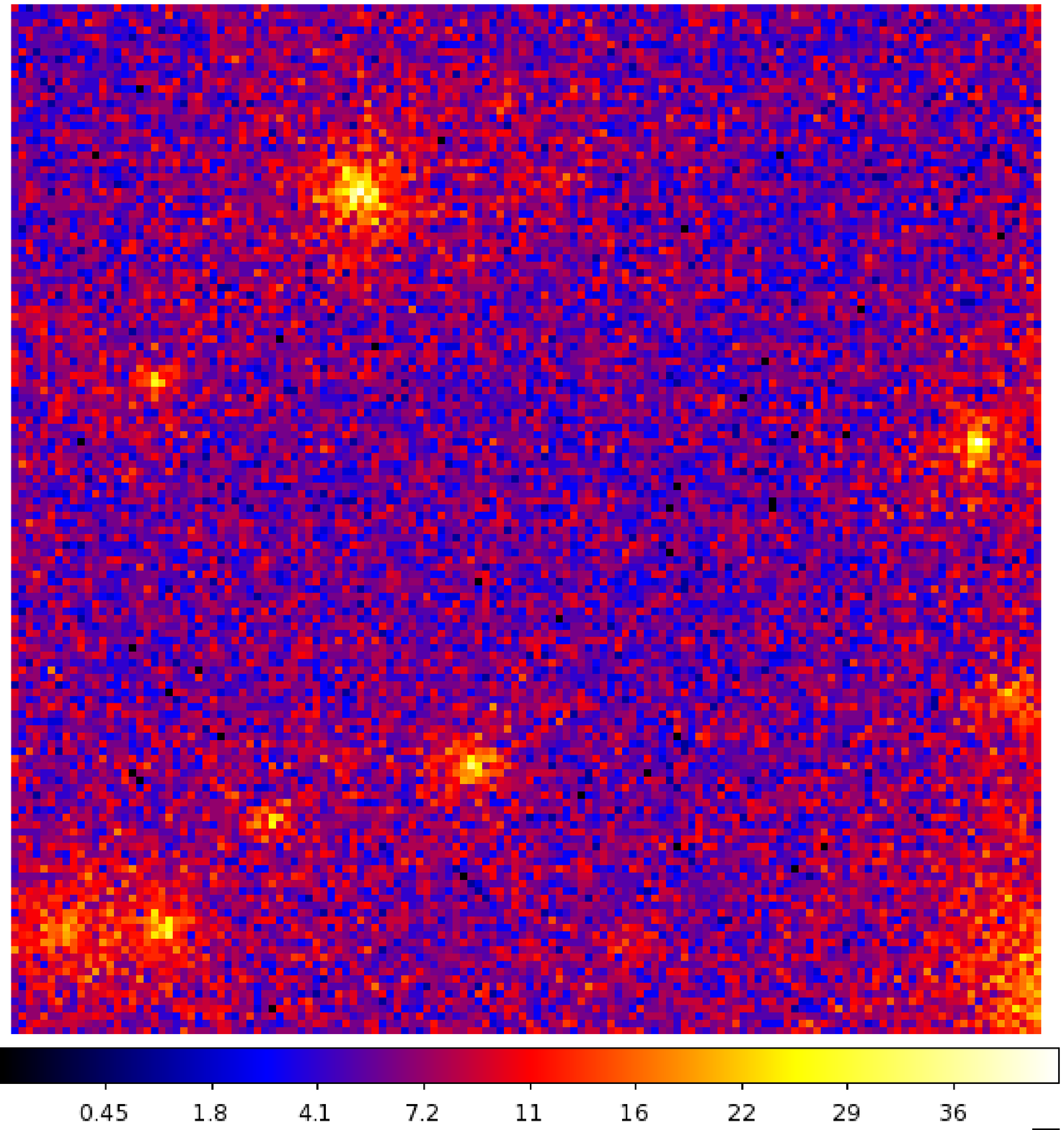}
\includegraphics[width=.33\textwidth]{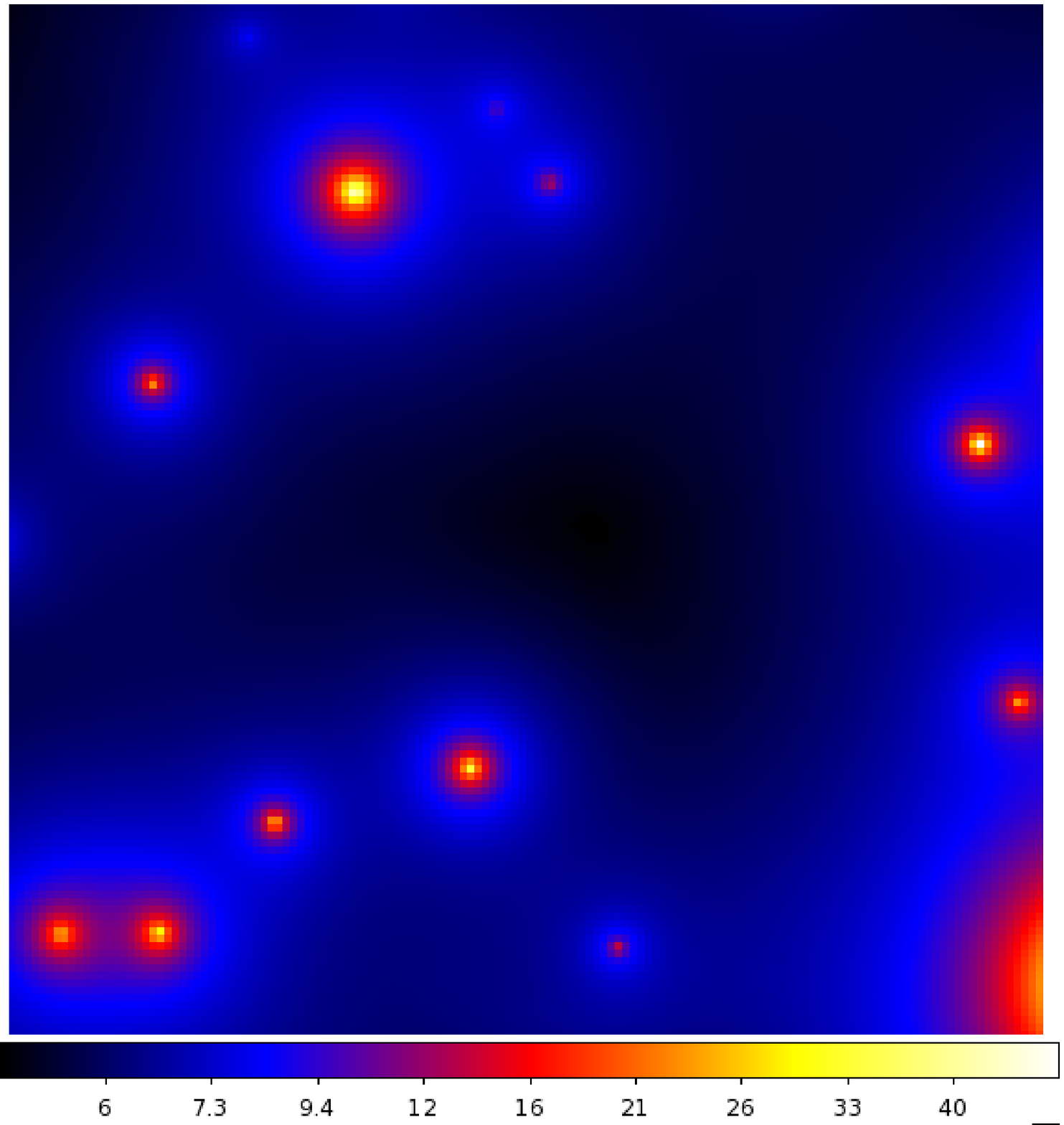}
\includegraphics[width=.33\textwidth]{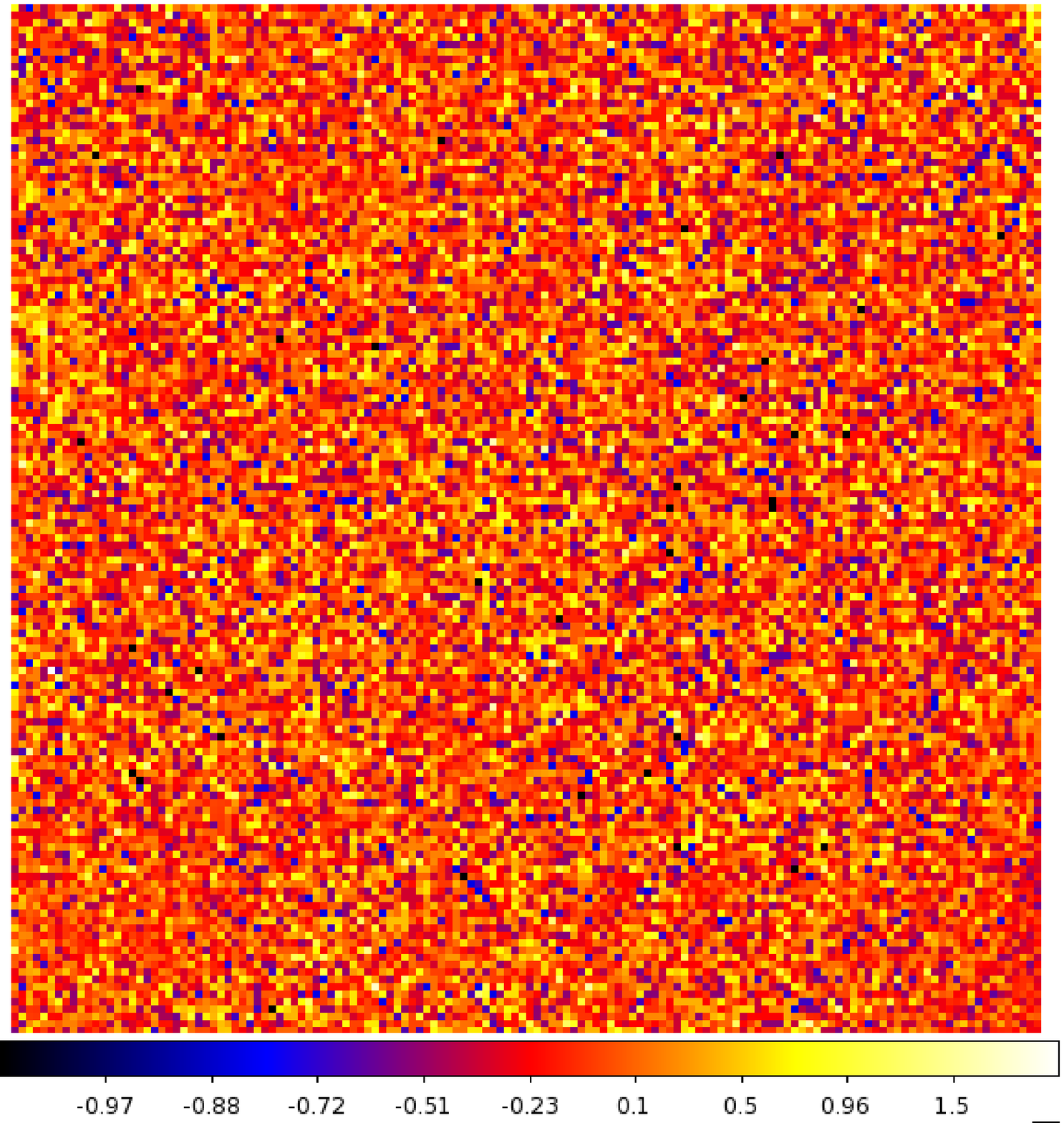}
\caption{\label{fig:photon} \emph{Left.} LAT photon count map for an 
area of $14\degr\times14\degr$ around the Coma galaxy cluster (whose
 center lies at the center of the image) obtained from about 5 years of
 observations. The cluster virial radius is about $1\fdg
 3$. \emph{Center.} Model count map for the basic analysis of the data
 with the 2FGL point sources, Galactic and extragalactic
 backgrounds. \emph{Right.} Residual
 map in percents obtained as $({\rm counts} - {\rm model})/{\rm
 model}$. All maps are in
 square-root scale for visualization purposes. }
\end{figure*}

We analysed 63~months (2008-08-04 15:43:37 to 2013-11-08 03:01:59)
of \emph{Fermi}-LAT \emph{reprocessed} Pass~7 (P7REP) data of the Coma galaxy cluster using 
\emph{Fermi Science Tools} (v9r32p5).\footnote{http://fermi.gsfc.nasa.gov/ssc/data/analysis/}
We adopted the standard event selection cuts suggested by the LAT collaboration
and analyzed events between 100~MeV and 100~GeV.
We analyzed both SOURCE and CLEAN events (\emph{Fermi} event class 
2 and 3, respectively), adopting the corresponding latest instrumental
response functions (P7REP$\_$V15), and we found good agreement between
the two. In the following, we report the results for the standard \emph{binned}
likelihood analysis of the SOURCE events only.

We select a region of interest (ROI) of $10\degr$ of radius around the
Coma cluster center ($\rmn{RA}=194\fdg 95$, $\rmn{DEC}=27\fdg 94$).
We then bin the data into $0\fdg 1$ pixels and 30 logarithmic steps in
energy.
This angular size is chosen to match the size of the point spread
function achieved by \emph{Fermi}-LAT at the highest energies, while it
is significantly worse at low energies (about $10\degr$ and $1\degr$ at
100 MeV and 1 GeV, respectively;  \citealp{2012ApJS..203....4A}).
The left panel of Figure~\ref{fig:photon} shows the photon count map of 
$14\degr\times14\degr$, used for our binned analysis.
The corresponding exposure map is computed for an area of
$40\degr\times40\degr$ centered on the cluster.

As a first step, we performed the analysis including all 26 point
sources within $15\degr$ from the cluster center, found in the 2-year
\emph{Fermi} catalog (2FGL; \citealp{2012ApJS..199...31N}) plus the
latest templates for the Galactic and extragalactic backgrounds provided
by the LAT collaboration (\emph{gll$\_$iem$\_$v05} and \emph{iso$\_$isource$\_$iv05}, respectively).
The point sources are modeled with the spectral characteristics given in
the 2FGL.
We allowed the spectral parameters of the 22 sources within the ROI to
vary in the likelihood analysis as well as the normalization of the
diffuse background components.
Note that no point sources are found within $3\degr$ from the cluster
center in the 2FGL.
The parameters of the other sources are kept fixed to the 2FGL values.

This model turns out to describe nicely the selected data, without the
need of adding any more point sources or diffuse components.
In Figure~\ref{fig:photon}, we show an image of the best-fit model
(center) as well as the residual map (right).
The latter is obtained by subtracting the model from the photon count
map in the left panel of Figure~\ref{fig:photon}, and then by dividing
again by the model.
It is a residual map in units of percent and fluctuates between
$-1$\% and $2$\%, and as shown below, is consistent with random fluctuations.

The cluster models that we test feature diffuse gamma-ray emission on a
very large scale and with quite low surface brightness.
Such kind of emission could be buried in the background emission, and
hence, a proper analysis is needed in order to draw conclusions.
Therefore, we run separate likelihood analysis with models including the
point-sources, Galactic and extragalactic backgrounds, in addition to a
given diffuse template for the cluster emission.
Those are described in detail in the next section.

Before to proceed with the diffuse templates, as a second step, we
placed an additional point-source (PS), modeled with a power-law
$E^{\Gamma}$, at the Coma center. We performed again the binned 
likelihood analysis with a fixed spectral index $\Gamma=-2$.
We find that the test statistics ($TS$; \citealp{TS_stat}) significance for 
this central point source is 0.\footnote{Note that in the background-only 
case, the $TS$ value can be converted to the usual definition of significance as 
$\sqrt{TS} \sigma$ (e.g., \citealp{2013arXiv1308.5654T}).}
We summarize the fit results, together with the obtained upper limits
(ULs), in Table~\ref{tab:results}.

In order to compare with the latest constraints obtained from 
\cite{2013arXiv1308.5654T}, we calculate the UL for their extended 
model (which corresponds to our PP model; see next section) for energies 
above 500~MeV obtaining $3.2 \times 10^{-10}$~cm$^{-2}$~s$^{-1}$. 
\cite{2013arXiv1308.5654T} obtained $4 \times 10^{-10}$~cm$^{-2}$~s$^{-1}$. 
Note that we adopt slightly different radius, mass and gas density values for the 
cluster modeling with respect to \cite{2013arXiv1308.5654T} which imply that
our total flux above 500~MeV is a factor of about 1.1 larger. We achieve a more 
stringent UL due to this choice and thanks to the longer observation time (they 
used 48~months of data). Note also that while \cite{2013arXiv1308.5654T}
uses CLEAN events, we report the result for the SOURCE events.

%%%%%%%%%%%%%%%%%%%%%%%%%%%%%%%%%%%%%%%%%%%%%%%%%%%%%%%%%%%%%%%%%%%
%%%%%%%%%%%%%%%%%%%%%%%%%%%%%%%%%%%%%%%%%%%%%%%%%%%%%%%%%%%%%%%%%%%
\section{Diffuse Emission from Cosmic Rays}
\label{sec:3}

In this section, we describe in detail the tested diffuse emission models.
We show some relevant model templates used in our analysis in 
Fig.~\ref{fig:templates}.

%%%%%%%%%%%%%%%%%%%%%%%%
\subsection{Gamma Rays from Pion Decays}

\citet{2010MNRAS.409..449P}, hereafter PP, performed hydrodynamical
simulations of galaxy clusters considering, in particular, diffusive
shock acceleration at structure formation shocks.
They provided predictions for the gamma-ray emission from CR protons and
electrons, and showed that the emission coming from pion decays dominates over 
the IC emission of both primary and secondary electrons for gamma-rays with an 
energy above 100~MeV. They then provide a semi-analytical model for the
pion-decay-induced emission that depends on a given cluster
mass and ICM density. The integral gamma-ray flux above 
the energy $E$ can be expressed as follows:
\begin{equation}
F_{\gamma,\rmn{PP}}(>E) = A_{\rmn{p}} \lambda_{\gamma,\rmn{PP}}(>E) \int_{V} k_{\rmn{PP}}(R) dV \,,
\end{equation}
where $\lambda_{\gamma,\rmn{PP}}(>E)$ and $k_{\rmn{PP}}(R)$ contain the 
spectral and spatial information, respectively, and are given in PP. 
$A_{\rmn{p}}$ is a dimensionless scale parameter related to the
maximum CR \emph{proton} acceleration efficiency $\xi_{p}$ for diffusive 
shock acceleration, which is the maximum ratio of CR energy density that 
can be injected with respect to the total dissipated energy at the 
shock.\footnote{{ The CR proton acceleration efficiency attains
its maximum, $\xi_{p}$, for high Mach number shocks, while it is 
lower for lower Mach numbers. The exact Mach number dependence of the acceleration 
efficiency is very uncertain. In this work we use as reference the \cite{2010MNRAS.409..449P}
simulations that depend on the \cite{2007A&A...473...41E} model for diffusive shock acceleration 
for which $\xi_{p}$, in the case of interest for clusters, is reached for Mach numbers of about 3. 
We note, however, that more detailed models such as the \cite{2013ApJ...764...95K} model, in the 
context of \emph{non-linear} diffusive shock acceleration, shows a different dependence, and the 
acceleration efficiency saturates at higher Mach numbers with respect to \cite{2007A&A...473...41E}.
Because of the early saturation in the \cite{2007A&A...473...41E} model, our constraints on the efficiency 
could be regarded as conservative in the low Mach number regime where more refined models show 
lower efficiencies.}} $A_{\rmn{p}} =1 $ for $\xi_{\rmn{p}} = 0.5$, and decreases for smaller efficiencies
obeying a non-linear relation (PP). However, note that the relation $A_{\rmn{p}}$--$\xi_{\rmn{p}}$ 
is linear for CR protons with an energy $\gtrsim10$~GeV which corresponds to pion decay 
emission with energies $\gtrsim1$~GeV (A.~Pinzke, private communication). 

The PP predictions have already being challenged by recent gamma-ray
observations \citep{2010ApJ...710..634A, 2011arXiv1111.5544M,
2011arXiv1105.3240P,2012arXiv1207.6749H,2012...VERITAS,2013arXiv1308.5654T}, 
suggesting either that the maximum CR proton acceleration efficiently at shocks is significantly lower
than 50\%, an optimistic value adopted in simulations, or the presence of CR streaming 
and diffusion out of the cluster core \citep{2011A&A...527A..99E,
2013arXiv1303.4746W,2012arXiv1207.6410Z}.

We test the PP spatial and spectral semi-analytical model for the Coma cluster, 
where the cluster mass is taken from \cite{2002ApJ...567..716R} and the ICM radial 
profile from \cite{1992A&A...259L..31B}.
In the PP model, only about 5\% of the total emission is coming from
radii beyond the virial radius, and therefore we decide to limit our
analysis within the virial radius\footnote{Defined with respect to a
density that is $200$ times the critical density of the Universe.}
$R_{200}=2.3$~$(h/0.7)^{-1}$\,Mpc.

Note, however, that assuming the magnetic field in the cluster is distributed according to
\begin{equation}
B(r) = B_{0} \left( \frac{\rho_{\rmn{gas}}(r)}{\rho_{\rmn{gas}}(0)} \right)^{\alpha_{\rmn{B}}}\,,
\end{equation}
where $\rho_{\rmn{gas}}$ is the ICM distribution, $B_{0} = 5$~$\mu$G and 
$\alpha_{\rmn{B}}=0.5$ as suggested by Faraday rotation (FR) measurements 
in Coma \citep{2010A&A...513A..30B}, the radio synchrotron emission 
(see, e.g., appendices of \citealp{2012arXiv1207.6410Z}) predicted by
the PP semi-analytical model does not match the spatial profile of the giant radio 
halo of the cluster at 1.4~GHz \citep{1997A&A...321...55D}, 
being much more peaked, as shown in Figure~\ref{fig:SB}. Additionally, 
it overproduces the central radio emission 
for a maximum acceleration efficiency $\gtrsim5$\% (assuming a linear 
scaling of $A_{\rmn{p}}$ with $\xi_\rmn{p}$). 

\begin{figure}
\centering 
\includegraphics[width=.15\textwidth]{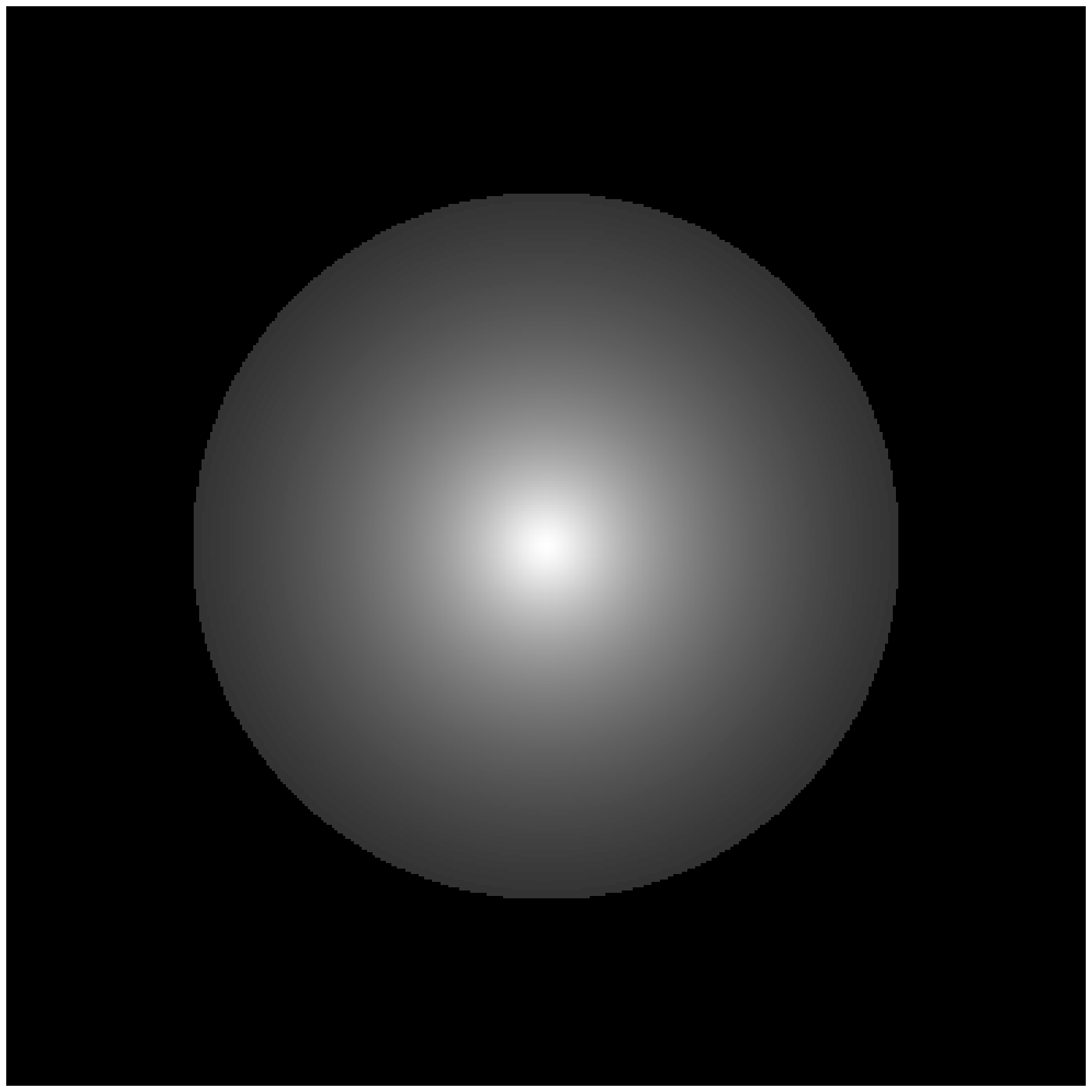}
\includegraphics[width=.15\textwidth]{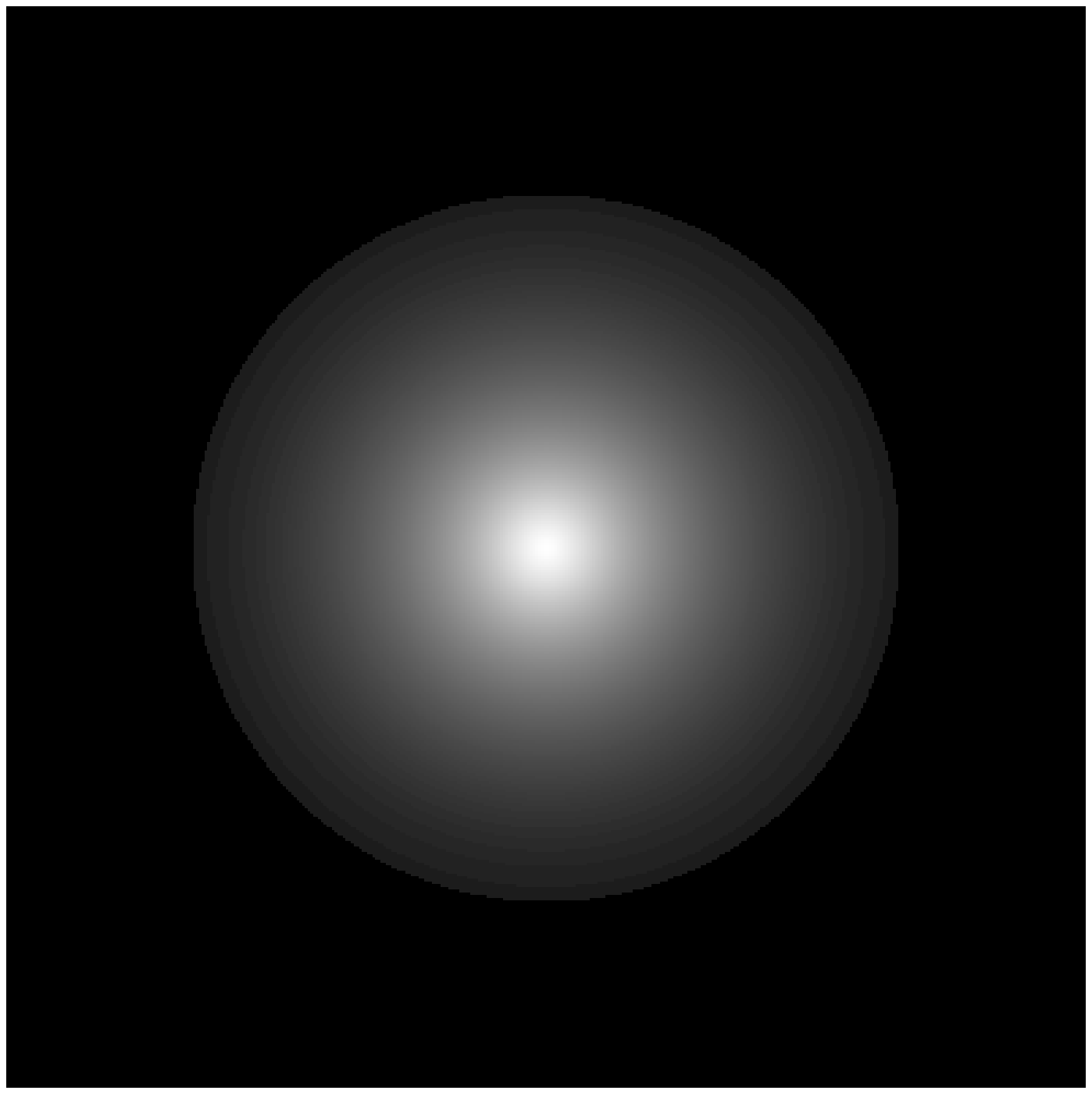}
\includegraphics[width=.15\textwidth]{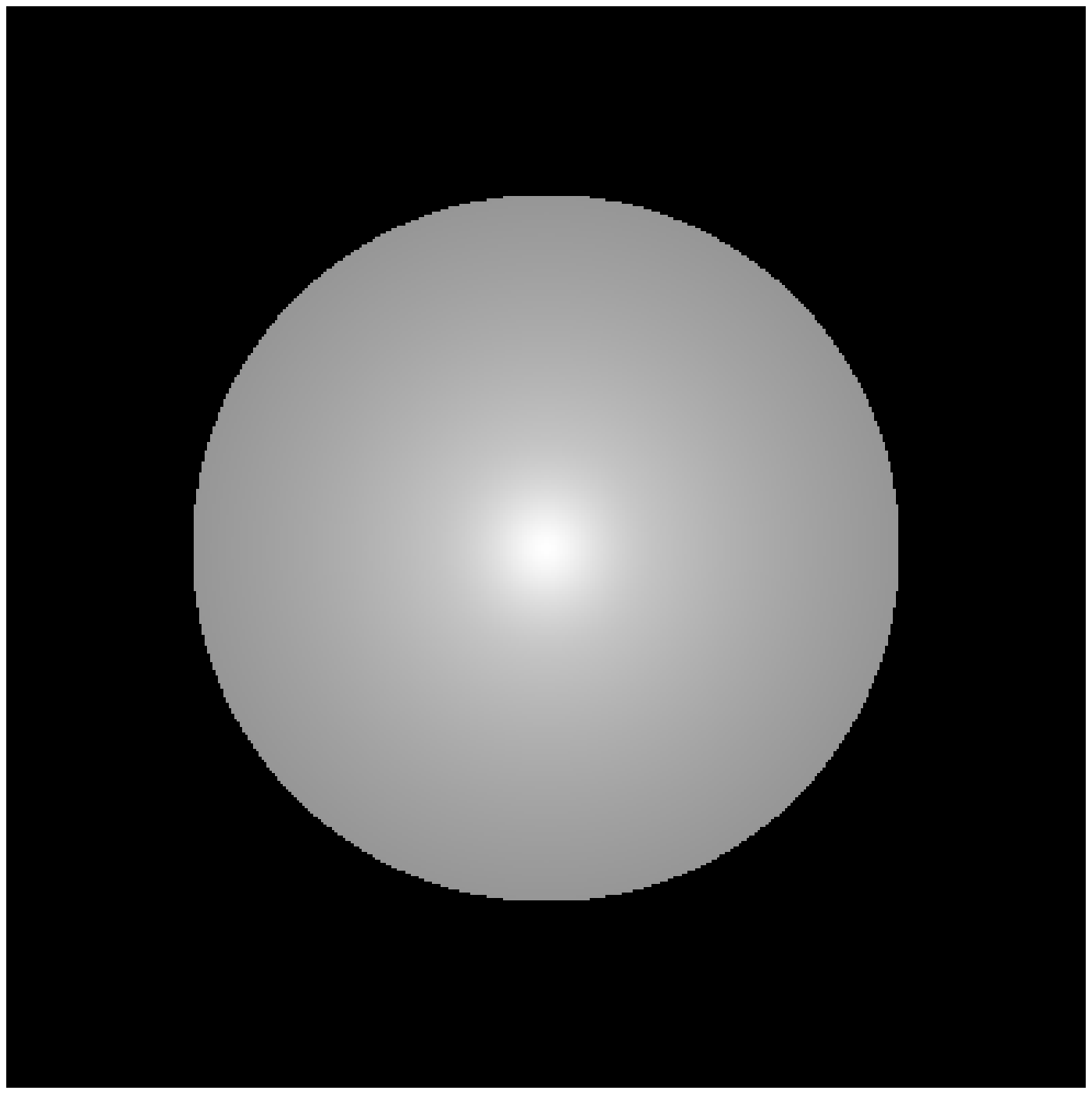}
\includegraphics[width=.15\textwidth]{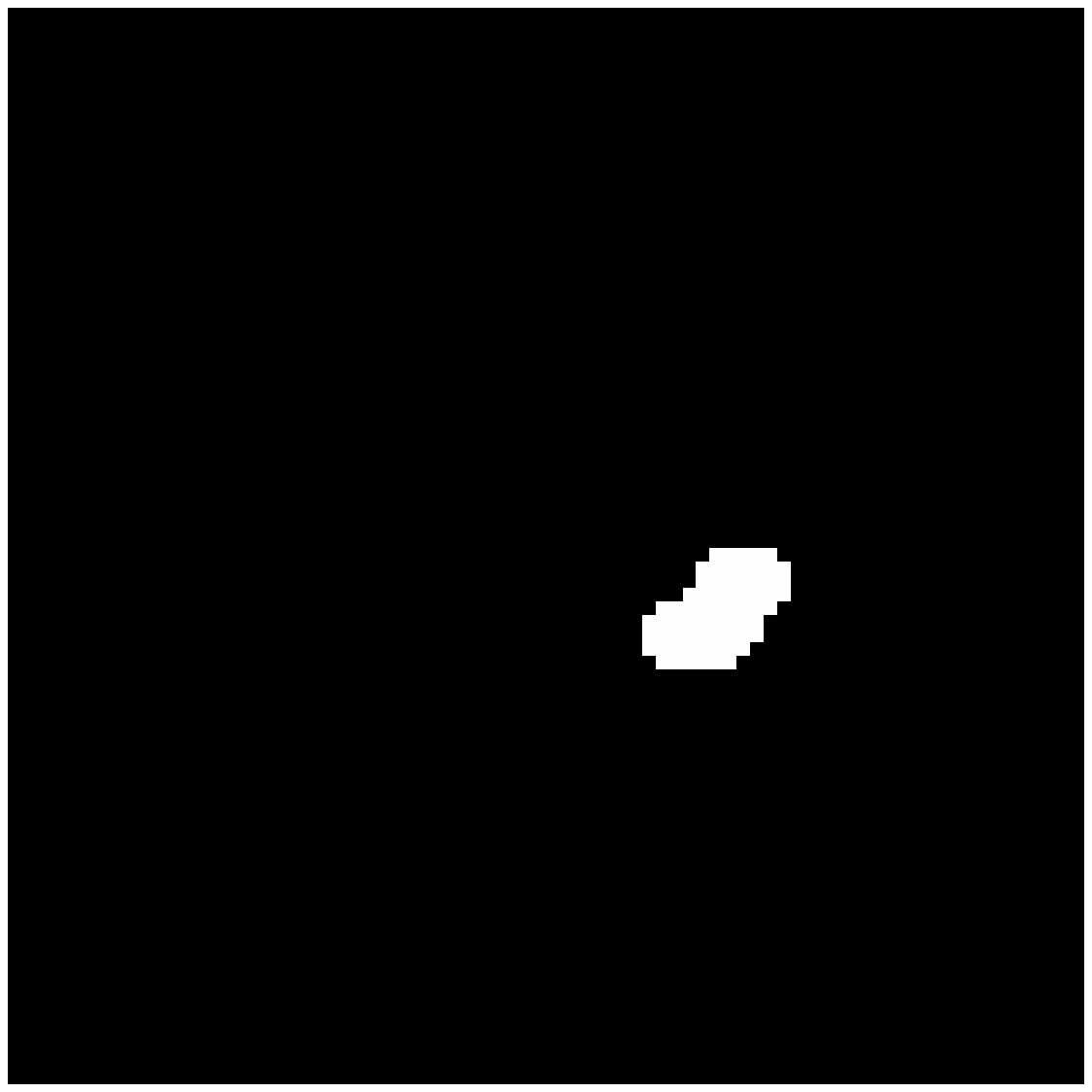}
\includegraphics[width=.15\textwidth]{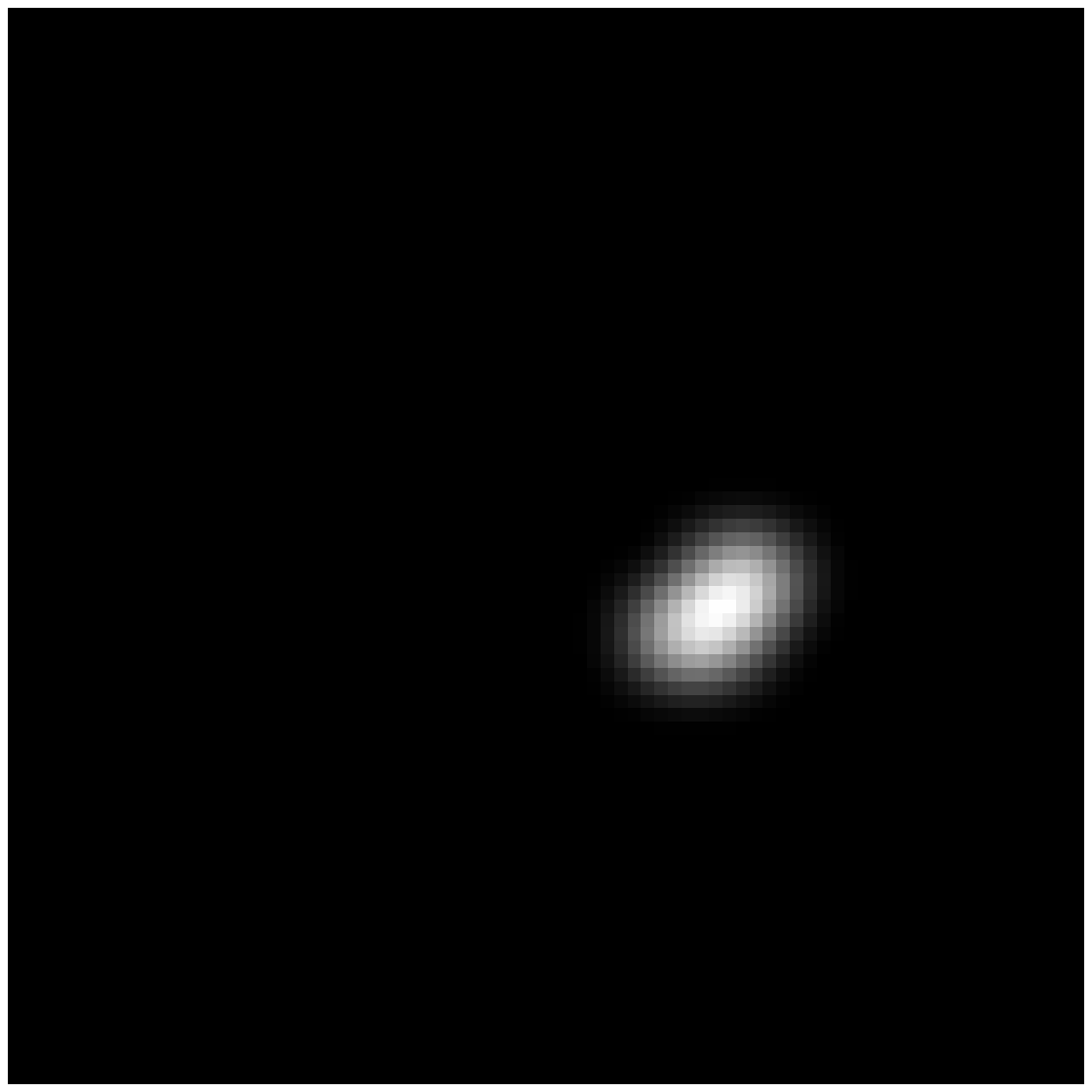}
\includegraphics[width=.15\textwidth]{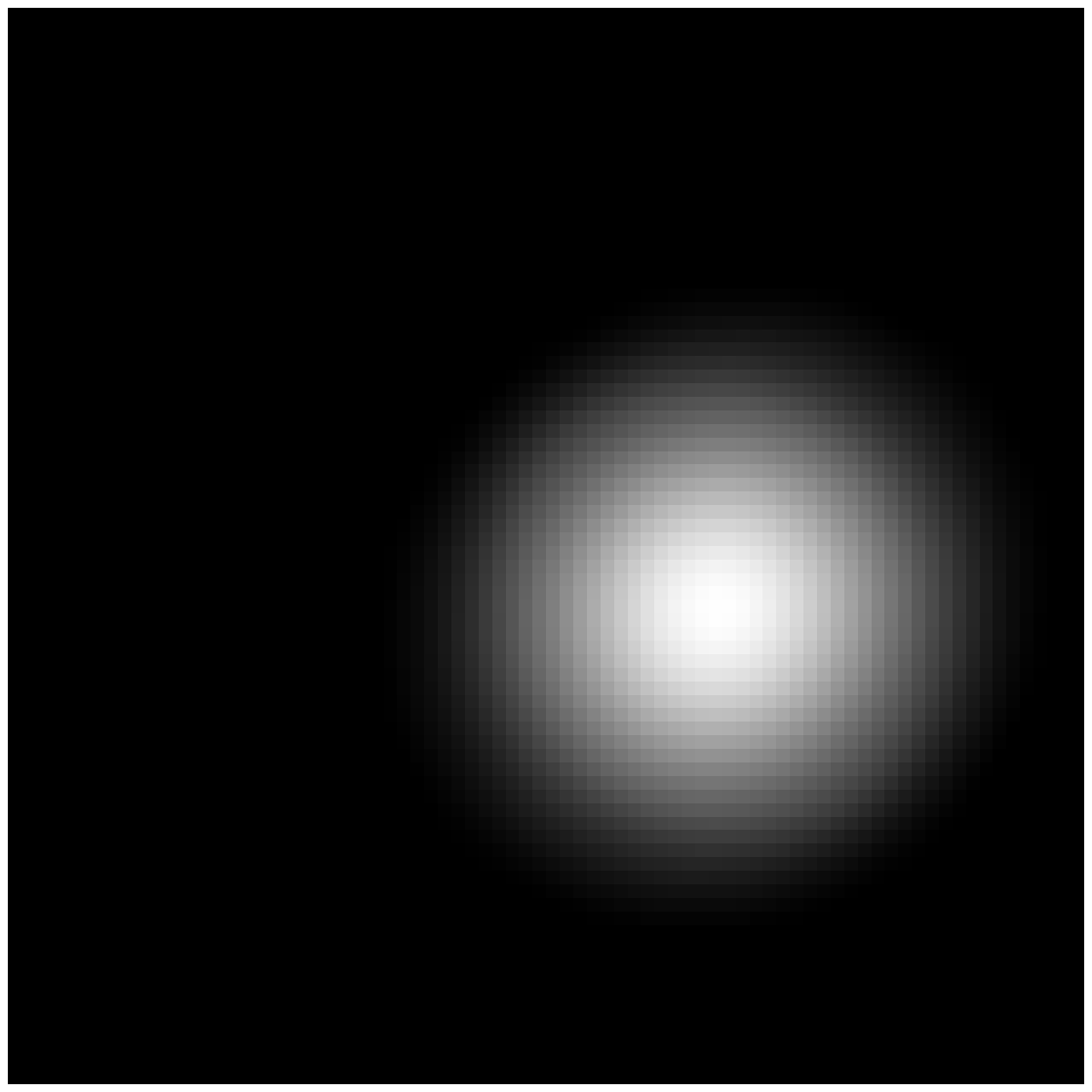}
\includegraphics[width=.15\textwidth]{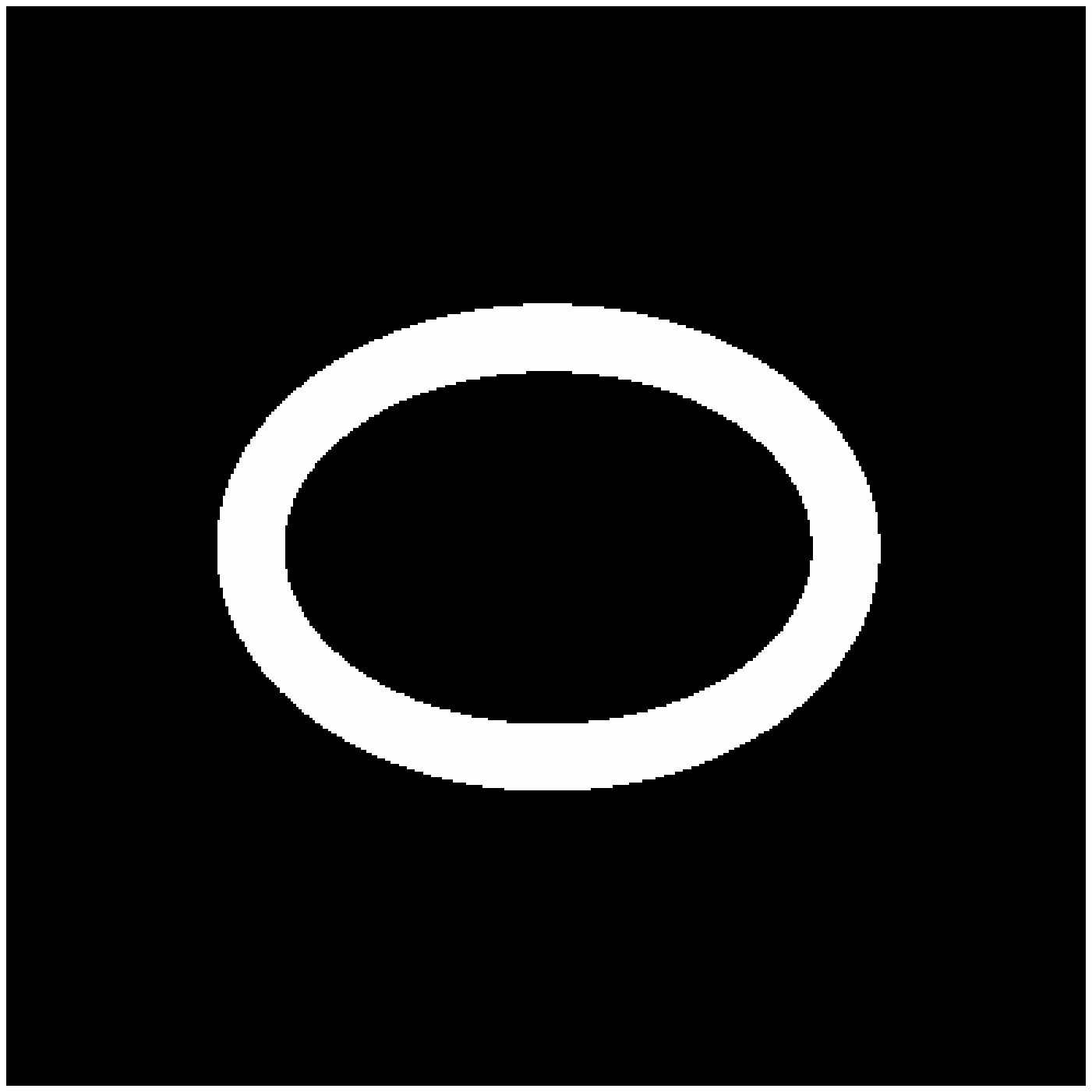}
\includegraphics[width=.15\textwidth]{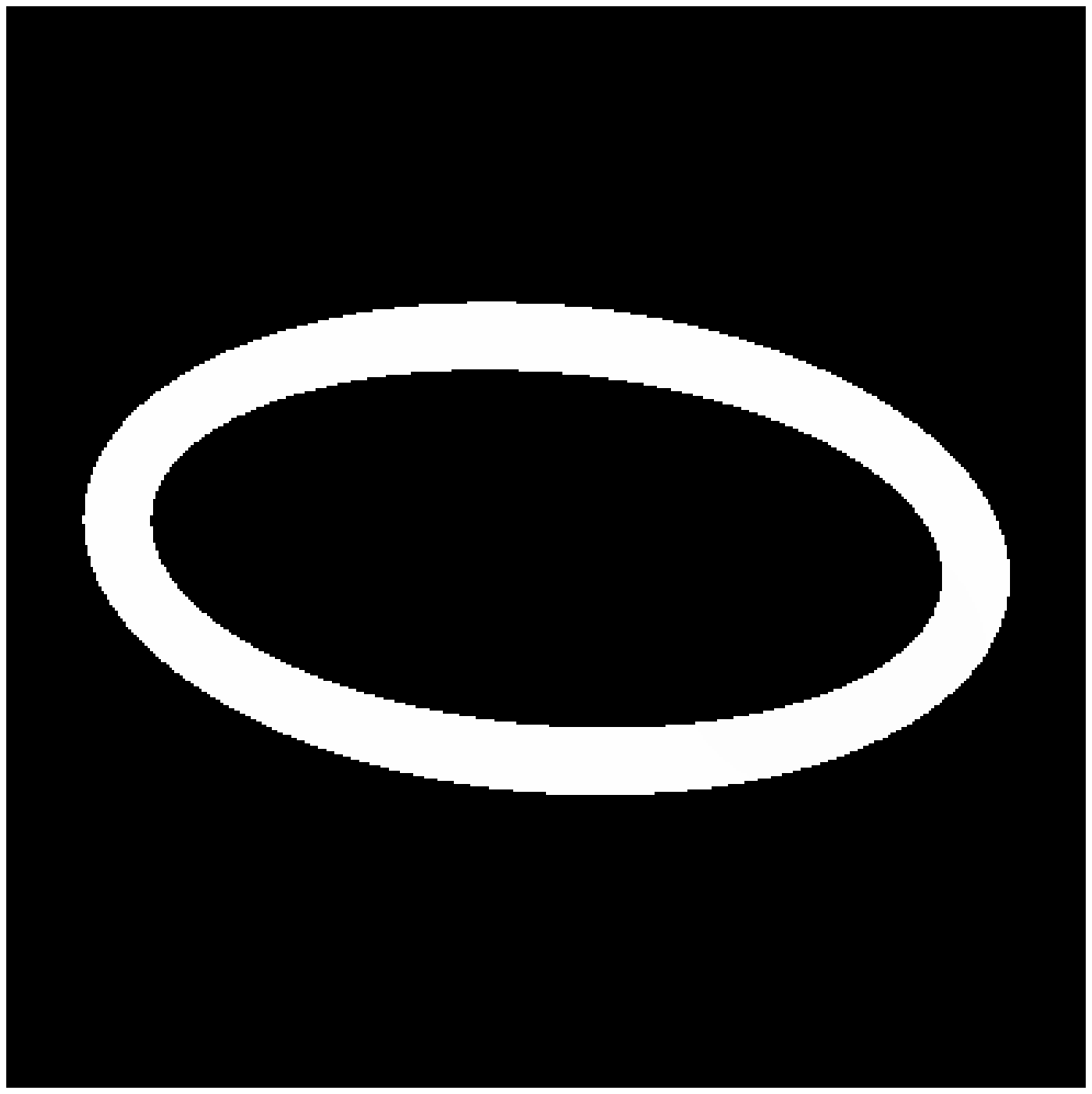}
\includegraphics[width=.15\textwidth]{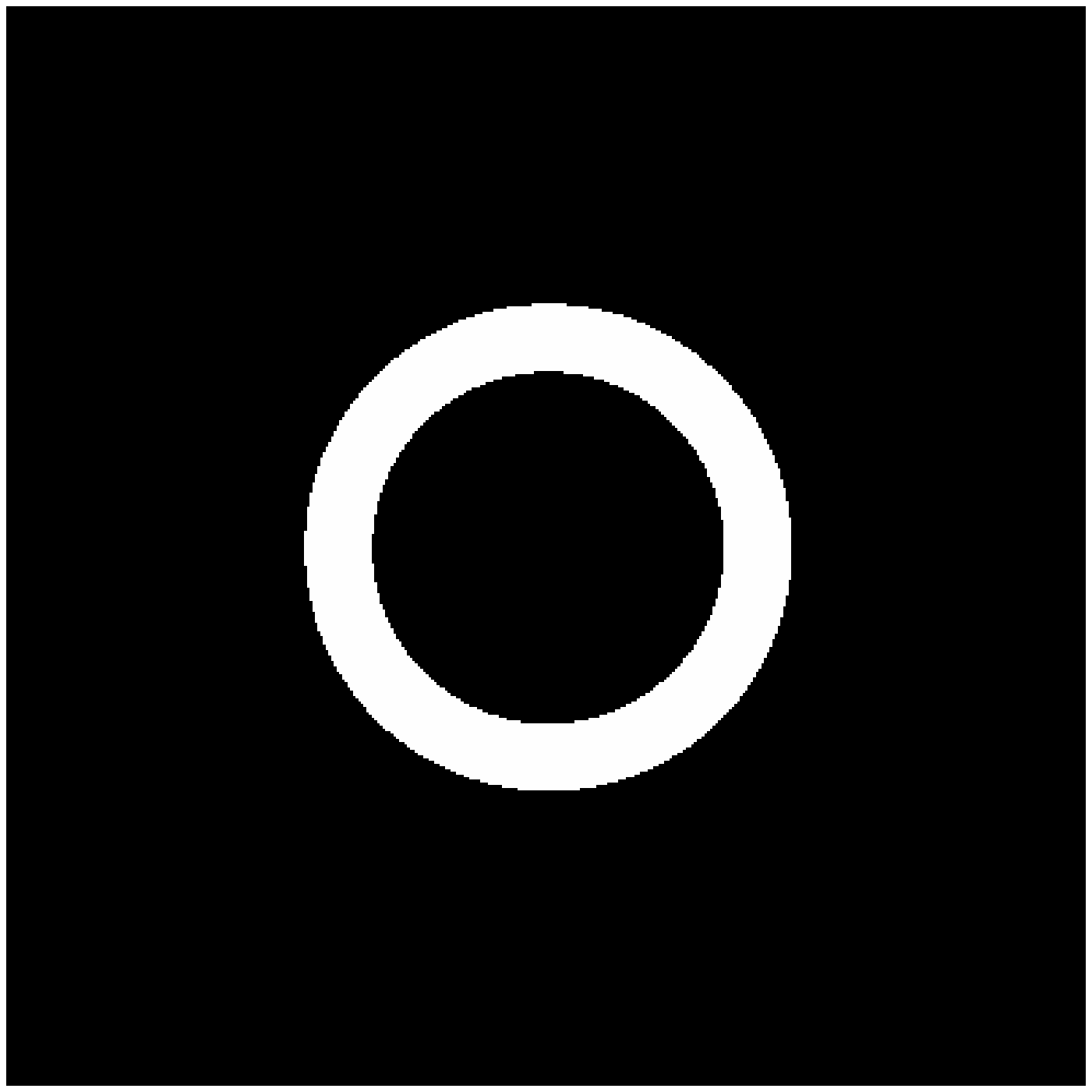}
\caption{\label{fig:templates} Some of the diffuse emission templates models 
used in the analysis. Top row shows, from left to right, $4\degr \times
4\degr$ images of the PP, ZPP-100 and ZPP-2 models, in logarithmic
scale. The middle row show the $8\degr \times 8\degr$ image of the
relic template, where the central and right images are after being
convolved with a Gaussian of width of $1\degr$ and $4\degr$,
respectively, to give an idea of the effect of the \emph{Fermi}-LAT
point-spread function at different energies. The bottom row shows,
from left to right, $8\degr \times 8\degr$ images of the ellipse,
tilted ellipse and ring models.}
\end{figure}

\cite{2012arXiv1207.6410Z}, hereafter ZPP, extended the PP
semi-analytic model with the inclusion of an effective
parameterization for CR transport phenomena, effectively redefining $k_{\rmn{PP}}(R)$ of eq.~(1).
Since the CR transport is determined by competition among advection due
to turbulent motion of gas, CR streaming, and diffusion, its efficiency
can be represented by a parameter,
\begin{equation}
 \gamma_{tu} = \frac{\tau_{st}}{\tau_{tu}},
\end{equation}
i.e., a ratio of a characteristic time scale of streaming, $\tau_{st}$,
and that of turbulence, $\tau_{tu}$ \citep{2011A&A...527A..99E}.
The parameter $\gamma_{tu}$ ranges from $100$, for highly turbulent
cluster and centrally peaked CR distributions, to $1$, for relaxed
clusters and flat distributions as CRs move toward the outskirts.
When $\gamma_{tu} \geq 100$, the model reproduce the
advection-dominated case of \cite{2010MNRAS.409..449P}, where
CR transport treatment is not included.
We test the ZPP model for the case of $\gamma_{tu} = 2$ (ZPP-2), 
matching the observed surface brightness profile of the Coma radio 
halo at 1.4~GHz (see Figure~\ref{fig:SB}).
Note that Coma is classified as merging cluster and one would expect it to 
be turbulent and not relaxed. Therefore, according to \citet{2011A&A...527A..99E}, 
high $\gamma_{tu}$ values and a centrally peaked CR distribution should be 
realized. \citet{2013arXiv1303.4746W} found a solution to this problem showing that,
when considering turbulent damping, turbulence may promote outward streaming 
more than inward advection, therefor allowing for flat CR profiles also in turbulent
clusters.

\begin{figure*}
\centering 
\includegraphics[width=.48\textwidth]{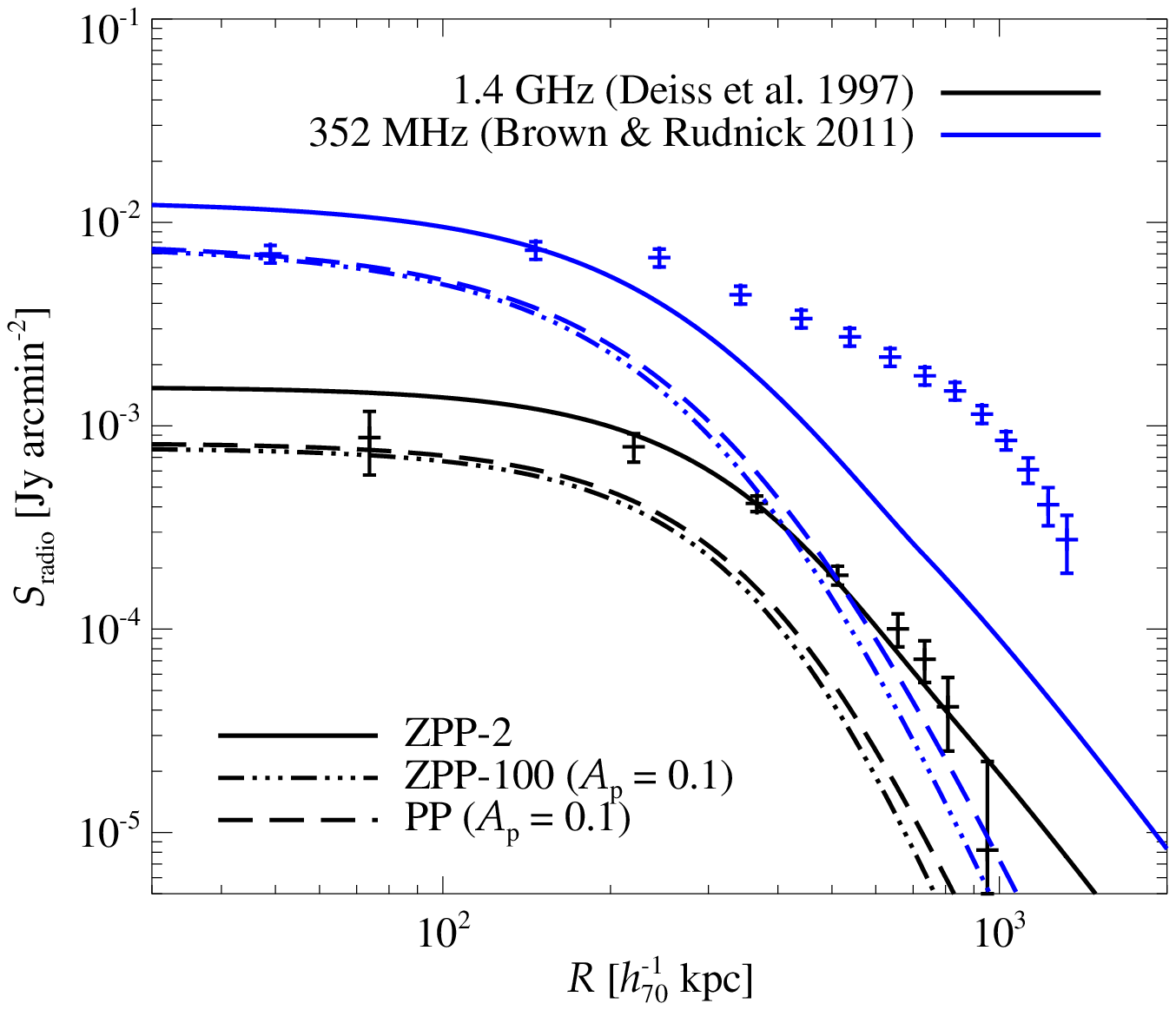}
\includegraphics[width=.48\textwidth]{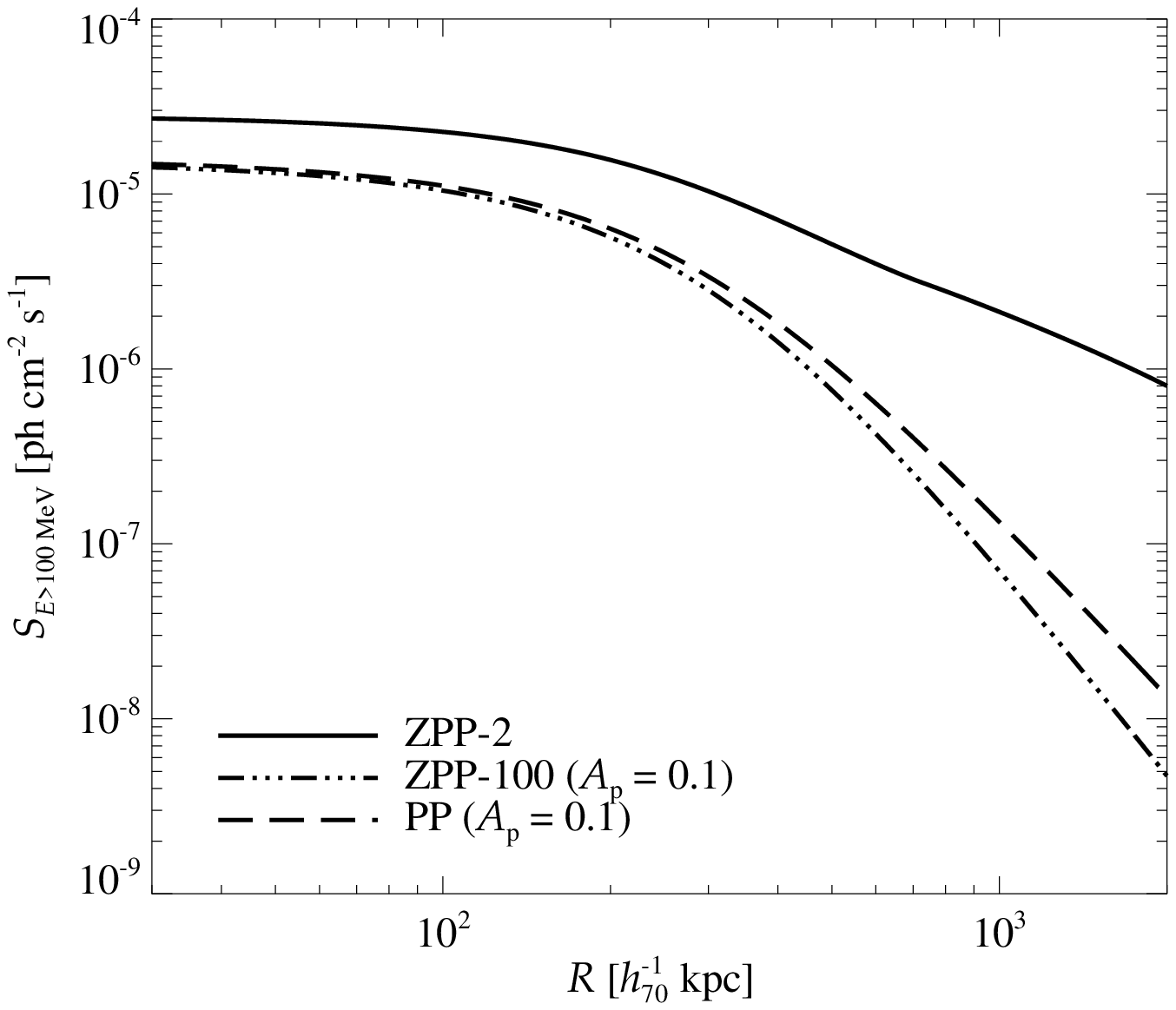}
\caption{\label{fig:SB} Surface brightness of the Coma cluster calculated as in ZPP. 
\emph{Left.} Surface brightness of the giant radio halo at 352~MHz \protect\citep{2011MNRAS.412....2B} 
and 1.4~GHz \protect\citep{1997A&A...321...55D}. Shown are the ZPP-2 model, and the
ZPP-100 and PP models scaled down with $A_{\rmn{p}}=0.1$ such that they do not overshot 
the radio emission. \emph{Right.} Gamma-ray surface brightness for energies above 100~MeV
for the ZPP-2 model, and ZPP-100 and PP models scaled down with $A_{\rmn{p}}=0.1$.}
\end{figure*}

At this point one may ask how representative is the giant radio halo of the Coma cluster, 
particularly the parameters' values needed for its modeling. While we cannot say with certainty
that these are common to all merging clusters hosting diffuse radio emission without an 
extensive analysis of the whole sample, we note that ZPP found the same characteristics,
in particular the need for low $\gamma_{\rmn{tu}}$ values, in the giant radio halo of the 
merging cluster Abell~2163. This is due to the large radial extension and shallow profile 
of the surface brightness of these objects, which appear to be a common property
among giant radio halos \citep{2012A&ARv..20...54F}.  

However, ZPP showed that even in the extreme case of a flat CR distribution, 
$\gamma_{tu}=1$, it is not possible to hadronically reproduce the $352$~MHz 
surface brightness of the giant radio halo of Coma \citep{2011MNRAS.412....2B}.
This favors re-acceleration models \citep{2012arXiv1207.3025B}, or
hybrid scenarios where only part of the radio emission is of hadronic
origin (see ZPP for an extensive discussion).
In this case, a centrally peaked CR distribution could still be
realized and only partially contribute to the total observed radio emission.
We therefore test also a ZPP model with $\gamma_{tu} = 100$ (ZPP-100).
Note that this is decreasing slightly faster toward the cluster
outskirts than the PP model because of the inclusion of the
characteristic radial decline of the temperature (ZPP; see
Figure~\ref{fig:templates} and \ref{fig:SB}).

Also for the ZPP models we limit our analysis within $R_{200}$. 
Both for PP and ZPP models, we let the normalization  of the emission to vary. 
The spectral shape is fixed to the PP model prediction, featuring the characteristic pion bump
at GeV energies followed by a concave spectrum that approaches
a power-law with spectral index of about 2.2 at TeV energies
(see Figure~12 of PP). We warn that by fixing the CR spectra to the 
PP findings, we exclude a potential free parameter that would affect our
conclusions \citep[see, e.g.,][]{2011arXiv1111.5544M,2012...VERITAS}.

%%%%%%%%%%%%%%%%%%%%%%%%
\subsection{Gamma Rays from Inverse-Compton Scattering}

\cite{2009JCAP...08..002K} developed an analytical model, adopting
a CR power-law energy spectrum with a spectral index of $-2$, and predicted 
that the IC emission from primary electrons accelerated at accretion shocks 
dominates over the pion-decay induced emission. Considering the differences
in acceleration efficiency and injected spectra may reconcile the findings by
\cite{2009JCAP...08..002K} and PP (see \citealp{2013arXiv1308.5654T},
for a detailed discussion). However, instead of the centrally concentrated 
gamma-ray emission from neutral pion decays, \cite{2009JCAP...08..002K}
found a spatially extended IC-induced emission out to the accretion shocks 
beyond the cluster's virial radius. This model predict a practically flat gamma-ray
emission up to the outer accretion shocks location where it should then peak
(see Fig.~2 of \citealp{2009JCAP...08..002K}). We therefore test this model
with a flat disk template of $1^\circ$ of radius. We consider a power-law spectrum 
with a fixed $\Gamma=-2$.
 
\cite{2009JCAP...08..002K} predict that the radio emission
should be dominated by secondaries. The corresponding profile of the radio 
surface brightness is similar to the ZPP model with a low $\gamma_{\rmn{tu}}$ value,
potentially being able to explain the 1.4~GHz data for the Coma giant halo but also suffering
the same problems discussed in the previous section when trying to reproduce the 352~MHz
data. Additionally, we note that \cite{2009JCAP...08..002K} assume a constant magnetic field in the 
cluster core, which is in contrast with FR estimates in Coma \citep{2010A&A...513A..30B},
and implies a shallower radio profile.
 
Accretion of intracluster gas should generate strong virial shock waves
around galaxy clusters and, according to several scenarios, this could potentially 
lead to ring-like emission features at different frequencies \citep{2000Natur.405..156L,
2000ApJ...545..572T,2000ApJ...545L..11W, 2002MNRAS.337..199M, 2003ApJ...585..128K,
2004ApJ...617..281K, 2005ApJ...623..632K, 2010JCAP...02..025K}. 
Note however that latest state-of-the-art cosmological hydrodynamical simulations
of galaxy clusters do not show the presence of such features
\citep{2008MNRAS.385.1211P,2008MNRAS.385.1242P,2010MNRAS.409..449P}.

Recently, \citet{2012arXiv1210.1574K} claimed the detection of a
ring-like structure in the gamma-ray sky map of the Coma cluster in the
VERITAS data \citep{2012...VERITAS} and interpreted this as IC-induced
emission due to accretion shocks around the cluster.
Therefore, we searched for the presence of this structure: it is a
$0\fdg 5$-wide elliptical shape with a semi-minor axis of about $1\fdg
3$, elongated toward the east-west direction.
We consider cases where the semi-major axis is of $2\degr$, and of $3\degr$
with the structure tilted towards the southwest direction for $5\degr$.
The latter case corresponds to the best fit of \cite{2012arXiv1210.1574K}, 
confirmed also by comparison with simulations by \cite{2003ApJ...585..128K}.
We refer to these as ellipse models.
We additionally consider the case of a $0\fdg 5$-wide ring with radius
of $R_{200}$ (ring model).
In all the cases, we uniformly fill the tested structure, and we
consider a power-law spectrum with a fixed $\Gamma=-2$.

We finally consider also a phenomenological template based on the 
radio relic observed in Coma outskirts, which is associated with
structure-formation phenomena.
Recent observations, both in radio and X-rays, support the idea that it
is connected to an infall shock front due to the NGC~4839 group falling
onto the cluster \citep{2011MNRAS.412....2B, 2013MNRAS.433.1701O,
2013arXiv1302.2907A, 2013arXiv1302.4140S} rather than a cluster-merger
shock.
Either way, the corresponding CR electrons could generate IC gamma-ray
emission.
We use the Green Bank Telescope (GBT) observation at 1.4~GHz presented
by \cite{2011MNRAS.412....2B} to generate a spatial template for the
relic.
We can use such an approach if we assume that the magnetic field is almost 
uniform across the relic \citep{2013MNRAS.tmp.1637B}.
This approximation is well justified in our case considering the poor
\emph{Fermi} angular resolution with respect to radio observations.
We therefore use the outermost contour of the GBT radio relic image from
Figure~2 of \citet{2011MNRAS.412....2B} to construct its spatial
template, which we uniformly fill.
We refer to this as the relic model.
We let the normalization to vary, and we use a power-law for the radio
spectrum with a fixed $\Gamma = -1.18$, as inferred from the 
observed radio spectrum \citep{2003A&A...397...53T}.

%%%%%%%%%%%%%%%%%%%%%%%%%%%%%%%%%%%%%%%%%%%%%%%%%%%%%%%%%%%%%%%%%%%
%%%%%%%%%%%%%%%%%%%%%%%%%%%%%%%%%%%%%%%%%%%%%%%%%%%%%%%%%%%%%%%%%%%
\section{Results and Implications}
\label{sec:4}

In the following subsections we discuss the implications of our findings for
each of the considered models. Table~\ref{tab:results} summarizes the
obtained ULs.

%%%%%%%%%%%%%%%%%%%%%%%%
\subsection{Pion Decay Emission}

In the PP and ZPP-100 models, assuming a maximum CR proton acceleration
efficiency of $50\%$, we would expect a total flux above 100~MeV and
within $R_{200}$ of $4.14$ and $3.24\times10^{-9}$~cm$^{-2}$~s$^{-1}$,
respectively.
The obtained ULs (see Table~\ref{tab:results}) are a factor of 0.26 and
0.28 of the theoretical expectations, respectively. 
This suggest that the maximum CR proton acceleration efficiency 
at shocks must be lower than about $15\%$, or implies the presence of 
significant CR propagation out of the cluster core in order to lower the 
central emission. Our ULs also set a limit to the CR-to-thermal pressure in 
the Coma cluster, $X_{\rmn{CR}} =P_{\rmn{CR}}/P_{\rmn{th}}$ 
(volume-averaged within $R_{200}$; see, e.g., ZPP), to be less then 
approximately 1.3\% and 0.6\% for the PP
and ZPP-100 model, respectively, within $R_{200}$.
Note that the limits on both the flux and CR pressure for the PP
model are more stringent than those obtained in the previous work on the
Coma cluster \citep{2011arXiv1105.3240P,
2012arXiv1207.6749H,2012...VERITAS,2013arXiv1308.5654T}.

If CR streaming and diffusion are in action in the cluster, we would
expect a much flatter emission profile which is represented by the ZPP-2
model.
This model matches the 1.4~GHz \cite{1997A&A...321...55D} surface
brightness profile of the Coma radio halo, assuming the magnetic field 
is distributed accordingly to FR measurements ($B_{0}=5~\mu$G, 
$\alpha_{\rmn{B}}=0.5$; \citealp{2010A&A...513A..30B}).
The predicted gamma-ray flux above 100~MeV within $R_{200}$ is
$2.36\times10^{-9}$~cm$^{-2}$~s$^{-1}$. In this case, $X_{CR}$
within $R_{200}$ is much higher, about 17\%, as streaming causes the
CR pressure to rise in the cluster's outskirts with respect to the
ICM pressure (see Figure~2 of ZPP). 
However, $X_{CR}$ reduces to 2.7\% within $R_{200}/2$. 
The corresponding flux UL shown in Table~\ref{tab:results},
$1.81\times10^{-9}$~cm$^{-2}$~s$^{-1}$, challenges the
ZPP-2 model. 
However, a slightly different choice of parameters can still circumvent
this limit while reproducing the 1.4~GHz radio data (ZPP), e.g., in case of 
$\gamma_{tu}=3$ and $\alpha_{\rmn{B}}=0.4$, the predicted gamma-ray 
flux above 100~MeV is about $1.3\times10^{-9}$~cm$^{-2}$~s$^{-1}$, 
whereas the UL hardly changes.

\begin{table}
\begin{center}
\caption{Results of the binned likelihood analyses for 63 months of the
 \emph{Fermi}-LAT data of the Coma cluster. The analyses include all 26 
 point sources within $15\degr$ from the cluster center, the
 extragalactic and galactic backgrounds, and a given model. All spectral
 templates are modeled as power law in the form of
 $\rmn{d}N/\rmn{d}E=N_{0}E^{\Gamma}$, except for PP and
 ZPP where the spectrum provided by the corresponding models is
 used. For each fit, reported are the resulting $TS$ significance,
 spectral index $\Gamma$ and flux UL $F_{\rmn{UL}}$ integrated 
 over 100~MeV--100~GeV with $95\%$ confidence level.}
%\medskip
\begin{tabular}{ccccc}
\hline\hline
\phantom{\Big|}
model  &  notes  &  $TS$  &  $\Gamma$  &  $F_{\rmn{UL}}$ \\
            &             &            &                      & [$\times 10^{-9}$~cm$^{-2}$~s$^{-1}$]  \\
\hline \\[-0.5em]
PS            &              &   0.0  &   $-2$                        & 0.62\\
PP            &              &  0.3  &     -                           & 1.08\\
ZPP-100   &  $\gamma_{tu}=100$   &  0.1  &    -    & 0.92\\
ZPP-2       &  $\gamma_{tu}=2$       &  1.3  &    -    & 1.81\\
Relic         &              &    0.0  &   $-1.18^{*}$                & 0.09\\
Ellipse      &              &    0.0  &  $-2$                      & 2.49\\
Ellipse      &  tilted    &    0.0  &  $-2$                       & 1.74\\
Ring         &              &    0.2  &  $-2$                      &  2.59\\
Disk         &  	       &    1.5  &  $-2$                       &  2.91\\
\hline
\end{tabular}
\label{tab:results}
\end{center}
\footnotesize{Notes. *The spectral index of the relic template is assumed to be 
as inferred from the observed radio spectrum \citep{2003A&A...397...53T}.
}
\end{table}

As explained above, an intriguing alternative is that of an hybrid scenario 
where the hadronic component would make up only the central part of the
observed radio emission (ZPP).
If this would be the case, the more centrally peaked PP and ZPP-100
models would still be a viable option, but requiring that they do not
over-shoot the radio emission both at 1.4~GHz and at 352~MHz.
Assuming $B_0 = 5$~$\mu$G and $\alpha_{\rmn{B}}=0.5$ sets both the PP
and ZPP-100 fluxes to be a factor of about 0.1 of the theoretical
expectations presented at the beginning of this section, corresponding
to a maximum CR proton acceleration efficiency of about $5$\%.
These are a factor of three lower than the \emph{Fermi}-LAT
ULs presented here.
However, there is a wide parameter space between a flat profile ($\gamma_{tu}
\sim 1$) and a totally advection-dominated profile ($\gamma_{tu} \sim 100$), leaving 
room for a possible detection of pion-decay emission in clusters with \emph{Fermi}-LAT 
or Cherenkov telescopes, in particular with the planned Cherenkov Telescope Array 
(CTA).\footnote{http://www.cta-observatory.org/}
Indeed, if such an hybrid scenario were realized in nature, the synergy
of radio and gamma-ray observations would be very important in
understanding the relevance of CR protons in clusters, and also in
breaking the degeneracy with magnetic field estimates and radio
modeling.

Another alternative is that the observed radio emission is not of hadronic
origin at all, but it is generated by re-acceleration of a seed population
of electrons. This seed population could be made of secondary
electrons, from CR proton-proton interactions with the ICM, that
are re-accelerated to emitting energies at a later stage
\citep{2012arXiv1207.3025B}. Also in this case, a corresponding
pion-decay induced emission is expected, but at a much lower 
level. Using the spectra shown in Figure~6 of \cite{2012arXiv1207.3025B}, 
we estimated that the integral gamma-ray flux for energies 100~MeV--100~GeV 
would be at a level of $2.6$--$1.4\times10^{-10}$~cm$^{-2}$~s$^{-1}$ in this
scenario, almost one order of magnitude lower than our current ULs.

%%%%%%%%%%%%%%%%%%%%%%%%
\subsection{Inverse-Compton Emission}
In this section we discuss the implications of our
analysis for the IC-induced emission from accretion
shocks, and connected to the radio relic of Coma.

\subsubsection{Accretion Shocks}
\cite{2009JCAP...08..002K} predicted a IC-induced flux above 100~MeV and within
$1^\circ$ from the center of the Coma cluster of about $10^{-8}$~cm$^{-2}$~s$^{-1}$.
Our UL obtained with the disk template is $2.9\times10^{-9}$~cm$^{-2}$~s$^{-1}$,
about a factor of 3 below their prediction. This limits the CR \emph{electron} acceleration
efficiency at shocks to be $\xi_{e}<1$\%.
The same is true for the prediction of \cite{2003ApJ...585..128K}. The absence of
any kind of ring-shaped emission around the Coma cluster, and the comparison
of our integral flux ULs above 100~MeV of about $2.5\times10^{-9}$~cm$^{-2}$~s$^{-1}$ 
with the predicted flux of about $4\times10^{-8}$~cm$^{-2}$~s$^{-1}$ 
(see eq.~(11) of \citealp{2012arXiv1210.1574K}), implies that the CR electron 
efficiency at shocks is $\xi_{e}<0.5$\%. This is consistent with \emph{Fermi} not having 
detected any of these structures \citep{2003ApJ...585..128K}.

\subsubsection{The Radio Relic}
\begin{figure}
\centering 
\includegraphics[width=.49\textwidth]{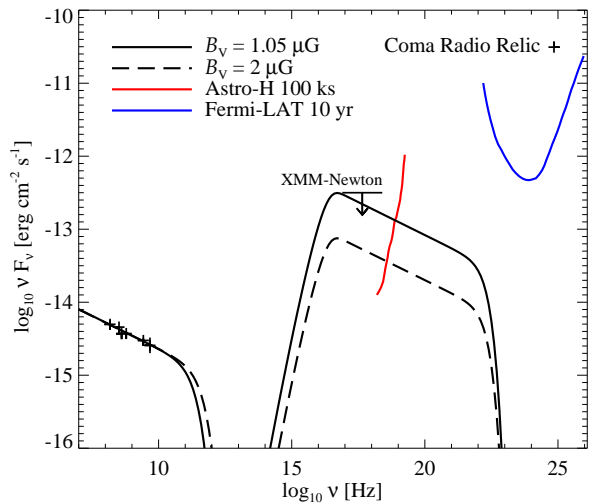}
\caption{\label{fig:synchIC} Spectral energy distribution of the radio relic of the Coma
cluster. The radio data is taken from the \protect\cite{2003A&A...397...53T} compilation,
while the \emph{XMM-Newton} UL is from \protect\cite{2006A&A...450L..21F}. We shown
the synchrotron and IC emission for $B_{\rmn{V}}=1.05$ and 2~$\mu$G. Also shown 
are the point-source sensitivities expected for Astro-H HXI in 100~ks \protect\cite{2012SPIE.8443E..1ZT} 
and for \emph{Fermi}-LAT in 10~yr \protect\citep{2013APh....43..348F}.}
\end{figure}

We model the emission of the radio relic of Coma using the data compiled by
\cite{2003A&A...397...53T}.
As explained above, the electrons generating the radio emission can also
emit hard X-rays and gamma-rays via IC scattering off the CMB photons.
We therefore compute their synchrotron and IC emission 
\citep{1970RvMP...42..237B, 1979rpa..book.....R}, as done in
\cite{2010A&A...514A..76M}.
We do not make any a priori assumption on the electron acceleration
mechanism, but simply adopt a phenomenological approach.
We assume a power-law electron distribution
$n(E) \propto E^{-\alpha_{\rmn{e}}}$.
The free parameters are the normalization of the electron 
distribution, spectral index $\alpha_{\rmn{e}}$, integration limits 
($E_{\rmn{min}}, E_{\rmn{max}}$), and the volume averaged magnetic field
$B_{\rmn{V}}$ across the relic region.
The electron spectral index is well determined by the slope of radio
spectrum $\alpha_{\nu}=1.18$ \citep{2003A&A...397...53T} and fixed at
$\alpha_{\rmn{e}}=2\alpha_{\nu}+1=3.36$.
The low- and high-energy cutoffs of the electron distribution are not
constrained by current data; in fact, the existing radio
measurements allow us only to establish $E_\rmn{min}/m_{e}c^{2} \lesssim
2500$ and $E_\rmn{max}/m_{e}c^{2} \gtrsim 2\times10^5$.
The low-energy cutoff is typically determined by the Coulomb losses and
we fix it at $E_\rmn{min}/m_{e}c^{2} = 200$ \citep[see, e.g.,][]{1999ApJ...520..529S}.
For the high-energy cutoff we assume $E_\rmn{max}/m_{e}c^{2} =
2\times10^5$, corresponding to electron energies of 100~GeV.
By varying the normalization such that the IC emission do not exceed
the high-energy ULs, we can estimate a lower limit for the magnetic
field needed to generate the observed synchrotron radio emission.
We found this to be constrained by the X-ray UL from \emph{XMM-Newton}
observations \citep{2006A&A...450L..21F}  as $B_{\rmn{V}} \gtrsim1$~$\mu$G. 
We show this in Figure~~\ref{fig:synchIC}. These values are consistent with the 
estimate of $B_{\rmn{V}}=2$~$\mu$G obtained from FR measurements in the cluster outskirts 
by \cite{2013MNRAS.tmp.1637B}. This is about a factor of 6 higher than what previously 
obtained by extrapolating eq.~(2) to the relic location \citep{2010A&A...513A..30B} and implies 
magnetic amplification in the relic region \citep{2013MNRAS.tmp.1637B}. In Figure~\ref{fig:synchIC}, 
we also show the case where we fix the magnetic field to $B_{\rmn{V}}=2$~$\mu$G. 

Not surprisingly, the X-ray upper limit obtained with $\emph{XMM-Newton}$ 
\citep{2006A&A...450L..21F} is much more constraining than \emph{Fermi}-LAT.
In fact, the \emph{Fermi}-LAT extragalactic (point-source) sensitivity for 10~years 
of observation \citep{2013APh....43..348F} is well above the expected IC flux.
Therefore, the high-energy emission associated with the electron population generating the 
Coma radio relic seems out of reach of existing and future-planned 
gamma-ray instruments. In this case, a much more exciting picture is that of the current and
next generations of X-ray satellites, such as NuSTAR (lunched in June 2012; \citealp{2013ApJ...770..103H}) 
and Astro-H (to be lunched in 2014; \citealp{2012SPIE.8443E..1ZT}). Indeed, 
the Astro-H Hard X-ray Imager (HXI) sensitivity curve (for a 100~ks exposure) shown 
in Figure~\ref{fig:synchIC} demonstrates potential to finally break the radio--X-ray degeneracy 
in determining the magnetic field value. We warn, however, that the radio relic emission
region\footnote{The emission modeled in Figure~\ref{fig:synchIC} 
corresponds to the radio relic region reported by \cite{2003A&A...397...53T}, which is about 
$800\times400$~kpc, corresponding to about $400$~arcmin$^2$. Note that the total relic extension
reported by \cite{2011MNRAS.412....2B}, on which we base out relic template for the analysis
of Section~2, is significantly larger with a transverse extent of about 2~Mpc. However, due to 
the very steep spectrum of the radio relic, we do not expect this to change our conclusion 
for the gamma-ray detectability.}
is about 0.1~deg$^2$, while the reported HXI sensitivity curve is for a point
source and the HXI angular resolution is below 2~arcmin \citep{2012SPIE.8443E..1ZT}. 

Note that the pion-decay emission should be suppressed at the relic due
to the lack of enough target ICM protons for hadronic interactions. 
Recently, \cite{2013arXiv1310.5707V} tested the acceleration of
protons by shock waves propagating through clusters against 
\emph{Fermi}-LAT ULs, finding, similarly to this work, that the
resulting central hadronic-induced emission lies very close to
them. Their prediction for the hadronic-induced gamma-ray emission 
\emph{at} the Coma relic, in the energy range of 200~MeV to 100~GeV, goes 
from $0.4$ to $3\times10^{-12}$~cm$^{-2}$~s$^{-1}$ depending on the
adopted model (F.~Vazza, private communication).\footnote{Note that
\cite{2013arXiv1310.5707V} assume that relics trace outward propagating 
shocks, and this is uncertain in the case of the Coma cluster as the 
relic may be tracing an infall shock \citep{2011MNRAS.412....2B, 2013MNRAS.433.1701O,
2013arXiv1302.2907A, 2013arXiv1302.4140S}.} 
This is very low, as expected from the low density of target protons at the relic location \citep{2013arXiv1302.4140S}. 
Our UL on the relic template in this energy range is around $0.7\times10^{-10}$~cm$^{-2}$~s$^{-1}$,
about two orders of magnitude above expectations.
 Note also that this UL is significantly
more stringent than the others we obtained because of the steeper spectrum.

We refrain from performing a similar synchrotron plus IC phenomenological modeling of the 
central giant radio halo of Coma as the corresponding volume average
magnetic field, of about $2$~$\mu$G \citep{2010A&A...513A..30B}, together with the 
extension of the radio emission area, of about $1$~deg$^2$, suggest that it would be 
extremely difficult to aim for a detection of the corresponding IC-induced emission, even 
with the next-generation X-ray satellites (see also \citealp{2011ApJ...727..119W}).

%%%%%%%%%%%%%%%%%%%%%%%%
\subsection{Diffuse Extragalactic Gamma-ray Background}
The constraints derived here not only affect particle acceleration 
modeling in clusters of galaxies, but also the possible contribution
to the extragalactic gamma-ray background (e.g., \citealp{2010PhRvL.104j1101A}). 

\cite{2000Natur.405..156L} estimated that the possible IC-induced emission
at accretion shocks could be high enough to entirely explain the gamma-ray 
background. \cite{2003ApJ...585..128K}, by using $N$-body simulations, estimated 
this to be about $10$\% for $\xi_{e}=5$\% (see also \citealp{2003APh....19..679G}). 
Our ULs imply that the possible IC-induced contribution must be lower than 1\%.

One can ask how much of the possible pion-decay-induced emission
in galaxy clusters could contribute to the extragalactic gamma-ray background.
\cite{Ando:2007yw}, with a simple analytical model, estimated this to be
less than a few percents (see also \citealp{1998APh.....9..227C}). By making 
use of the mock galaxy clusters catalogs of \cite{2013arXiv1311.4793Z},
\footnote{The \cite{2013arXiv1311.4793Z} mock catalogs 
have been taken from the MultiDark online database (www.multidark.org, 
\citealp{2011arXiv1109.0003R})} which include the prediction for the pion-decay
induced emission in galaxy clusters following the ZPP prescription, we estimate
this to be less than 1\%.

Summarizing, this means that acceleration of CR protons \emph{and} electrons 
in galaxy clusters gives a negligible contribution to the diffuse extragalactic 
gamma-ray background. We note, however, that if the other possible contributions,
such as from blazars and star-forming regions, are well understood, this could 
be a potentially interesting way to study relativistic particles in galaxy clusters.

%%%%%%%%%%%%%%%%%%%%%%%%%%%%%%%%%%%%%%%%%%%%%%%%%%%%%%%%%%%%%%%%%%%
%%%%%%%%%%%%%%%%%%%%%%%%%%%%%%%%%%%%%%%%%%%%%%%%%%%%%%%%%%%%%%%%%%%
\section{Conclusions}
\label{sec:6}
Since generation of shocks and particle acceleration in the shocks are
generic predictions of large-scale-structure formation scenarios,
one expects gamma-ray emission from clusters of galaxies by 
secondaries from CR proton-proton interactions with the ICM, 
and by IC scattering of CR electrons off the CMB. 
In this paper, we analyzed 63-month (P7REP) data from {\it Fermi}-LAT photons 
between 100~MeV and 100~GeV from the Coma cluster, one of the best
studied nearby galaxy clusters.
Coma also shows recent activity of a merger and accretion, and has both 
a radio relic and giant radio halo.
These observations make Coma a promising gamma-ray source, such that
one is able to test CR energy content in the galaxy cluster and also
particle acceleration mechanisms.
We tested several template models of the gamma-ray emission from Coma
and found no positive signature corresponding to any of these models.
We, however, obtained the most stringent constraints to date on the
Coma cluster above 100~MeV, as summarized below (see also Table~\ref{tab:results}).

(1) {\it Point-source model.} We obtained a point-source flux upper limit,
assuming a power-law energy spectrum with an index of $-2$, 
of $F_{\rm UL} = 0.6 \times 10^{-9}$~cm$^{-2}$~s$^{-1}$,
which is better by a factor of a few compared with previous
studies~\citep[e.g.,][]{2012JCAP...07..017A}. 
Note, however, that a softer spectrum would cause 
the UL to increase (as in \citealp{2013arXiv1308.5654T}).

(2) {\it Pion decays.} In this case, the spatial profile depends on the 
efficiency of CR turbulent motion compared with that of streaming.
We chose several profiles and found that the flux limits are $F_{\rm UL}
\simeq (0.9$--$1.8) \times 10^{-9}$~cm$^{-2}$~s$^{-1}$, where the latter
(former) value is for a more (less) extended model as a result of higher
(lower) efficiency of streaming. These limits constrain 
the predictions of Pinzke \& Pfrommer (2010) and 
\cite{2012arXiv1207.6410Z}, which implies 
either that maximum CR proton acceleration efficiencies at shocks are lower
than about 15\%, or the presence of significant CR propagation out
of the cluster core. We also note that by comparing the advection-dominated
centrally peaked profiles to the observed radio emission, the maximum CR
proton acceleration efficiency is limited to be below about 5\%. Note,
however, that these conclusions rely on the assumption of magnetic field 
estimates from FR measurements \citep{2010A&A...513A..30B} and on a 
fixed CR spectrum (PP). 

(3) {\it Inverse-Compton Emission.}
Motivated by predictions for IC-induced emission from electrons
accelerated at accretion shocks, we investigated both
a disk and ring-like emission template. We found ULs of $F_{\rm UL} =
(1.7$--$2.9) \times 10^{-9}$~cm$^{-2}$~s$^{-1}$, which is not consistent with low-energy
extrapolation of a recent claim of positive detection of such a ring-like feature 
in the VERITAS data~\citep{2012arXiv1210.1574K}. Additionally, this
limits the CR electron acceleration efficiency at shocks to be less
than 1\% both in the \cite{2003ApJ...585..128K} and in the 
\cite{2009JCAP...08..002K} scenarios.  

(3) {\it Radio Relic.}
We adopted an emission profile consistent with the Coma radio relic, and looked 
for the corresponding gamma-ray emission. The radio emission from the relic is 
interpreted as a synchrotron radiation from non-thermal electrons, and there 
should be a corresponding high-frequency component due to IC scattering
from the same electrons. The gamma-ray-flux upper limit is $F_{\rm UL} = 0.9 
\times 10^{-10}$~cm$^{-2}$~s$^{-1}$, but this is too weak to constrain the
electron population. This is because the expected energy range of the IC
scattering off CMB photons is in X-rays for an electron population matching
the radio relic synchrotron emission. Instead, we find that the current (NuSTAR) 
and future (Astro-H) X-ray telescopes have excellent prospects for detecting this
IC emission.  

(4) {\it Diffuse Extragalactic Gamma-ray Background.}
We conclude by noting that, following \cite{2003ApJ...585..128K} our results imply that 
the possible IC-induced emission associated with structure formation shocks in clusters 
of galaxies can contribute to the diffuse extragalactic gamma-ray background by less 
than 1\%. At the same time, using the \cite{2013arXiv1311.4793Z} mock
galaxy cluster catalogs, we estimate that also the possible pion-decay induced 
emission can contribute only by less than 1\%. This renders the contribution 
to the diffuse extragalactic gamma-ray background due to the high-energy photons 
from structure-formation processes in clusters of galaxies negligible. 

%%%%%%%%%%%%%%%%%%%%%%%%%%%%%%%%%%%%%%%%%%%%%%%%%%%%%%%%%%%%%%%%%%%
%%%%%%%%%%%%%%%%%%%%%%%%%%%%%%%%%%%%%%%%%%%%%%%%%%%%%%%%%%%%%%%%%%%
\section*{Acknowledgments}
We would like to thank Anders Pinzke and Christoph Weniger for useful discussions.
We thank Franco Vazza for providing the pion-decay emission at the radio relic
location as given by his model.
FZ~thank the support of the Spanish MICINN's Consolider-Ingenio 2010 Programme 
under grant MultiDark CSD2009-00064, AYA10-21231.
The MultiDark Database used in this paper and the web application providing online 
access to it were constructed as part of the activities of the German Astrophysical Virtual 
Observatory as result of a collaboration between the Leibniz-Institute for Astrophysics 
Potsdam (AIP) and the Spanish MultiDark Consolider Project CSD2009-00064. 
The Bolshoi and MultiDark simulations were run on the NASA's Pleiades supercomputer 
at the NASA Ames Research Center. The MultiDark-Planck (MDPL) and the BigMD simulation 
suite have been performed in the Supermuc supercomputer at LRZ using time granted by PRACE.
This work was supported by the Netherlands Organization for Scientific
Research (NWO) through Vidi grant.

%%%%%%%%%%%%%%%%%%%%%%%%%%%%%%%%%%%%%%%%%%%%%%%%%%%%%%%%%%%%%%%%%%%
%%%%%%%%%%%%%%%%%%%%%%%%%%%%%%%%%%%%%%%%%%%%%%%%%%%%%%%%%%%%%%%%%%%
\bibliographystyle{mn2e}
\bibliography{bib_file}

%%%%%%%%%%%%%%%%%%%%%%%%%%%%%%%%%%%%%%%%%%%%%%%%%%%%%%%%%%%%%%%%%%
%%%%%%%%%%%%%%%%%%%%%%%%%%%%%%%%%%%%%%%%%%%%%%%%%%%%%%%%%%%%%%%%%%
\label{lastpage}

\end{document}